\documentclass[reprint,superscriptaddress,showpacs,
amsmath,amssymb,aps,floatfix,prl]{revtex4-1}

\usepackage{graphicx}
\usepackage{dcolumn}
\usepackage{bm}

\begin{document}

\title{Emergent high-spin state above 7~GPa in superconducting FeSe}% 

\author{B.~W.~Lebert}
\affiliation{IMPMC-Sorbonne Universit\'{e}s, Universit\'{e} Pierre et Marie Curie, CNRS, IRD, MNHN 4, place Jussieu, 75252 Paris, France}
\affiliation{Synchrotron SOLEIL, L'Orme des Merisiers, BP 48 St Aubin, 91192 Gif sur Yvette}

\author{V.~Bal\'edent}
\affiliation{Laboratoire de Physique des Solides, CNRS,  Univ. Paris-Sud, Universit\'e Paris-Saclay 91405 Orsay cedex, France}

\author{P.~Toulemonde}
\affiliation{CNRS, Institut N\'{e}el, F-38000 Grenoble, France}
\affiliation{Universit\'{e} Grenoble-Alpes, Institut N\'{e}el, F-38000 Grenoble, France}

\author{J.~M.~Ablett}
\affiliation{Synchrotron SOLEIL, L'Orme des Merisiers, BP 48 St Aubin, 91192 Gif sur Yvette}

%\author{S.~Klotz}
%\affiliation{IMPMC-Sorbonne Universit\'{e}s, Universit\'{e} Pierre et Marie Curie, CNRS, IRD, MNHN 4, place Jussieu, 75252 Paris, France}

%\author{T.~Hansen}
%\affiliation{Institut Laue-Langevin, B.P. 156, 38042 Grenoble Cedex 9, France}

%\author{P.~Rodi\`{e}re}
%\affiliation{CNRS, Institut N\'{e}el, F-38000 Grenoble, France}
%\affiliation{Universit\'{e} Grenoble-Alpes, Institut N\'{e}el, F-38000 Grenoble, France}

%\author{M.~Raba}
%\affiliation{CNRS, Institut N\'{e}el, F-38000 Grenoble, France}
%\affiliation{Universit\'{e} Grenoble-Alpes, Institut N\'{e}el, F-38000 Grenoble, France}

\author{J.-P.~Rueff}
\affiliation{Synchrotron SOLEIL, L'Orme des Merisiers, BP 48 St Aubin, 91192 Gif sur Yvette}
\affiliation{LCPMR-Sorbonne Universit\'{e}s, Universit\'{e} Pierre et Marie Curie, CNRS, 4 place Jussieu, 75252 Paris, France}
\date{\today}
  
\begin{abstract}
The local electronic and magnetic properties of superconducting FeSe have been investigated by K$\beta$ x-ray emission (XES) and simultaneous x-ray absorption spectroscopy (XAS) at the Fe K-edge at high pressure and low temperature. Our results indicate a sluggish decrease of the local Fe spin moment under pressure up to 7~GPa, in line with previous reports, followed by a sudden increase at higher pressure which has been hitherto unobserved. The magnetic surge is preceded by an abrupt change of the Fe local structure as observed by the decrease of the XAS pre-edge region intensity and corroborated by ab-initio simulations. This pressure corresponds to a structural transition, previously detected by x-ray diffraction, from the $Cmma$ form to the denser $Pbnm$ form with octahedral coordination of iron. Finally, the near-edge region of the XAS spectra shows a change before this transition at 5~GPa, corresponding well with the onset pressure of the previously observed enhancement of $T_c$. Our results emphasize the delicate interplay between structural, magnetic, and superconducting properties in FeSe under pressure.
\end{abstract}


\maketitle
\begin{figure}[thbp]
\includegraphics[width=3.375in]{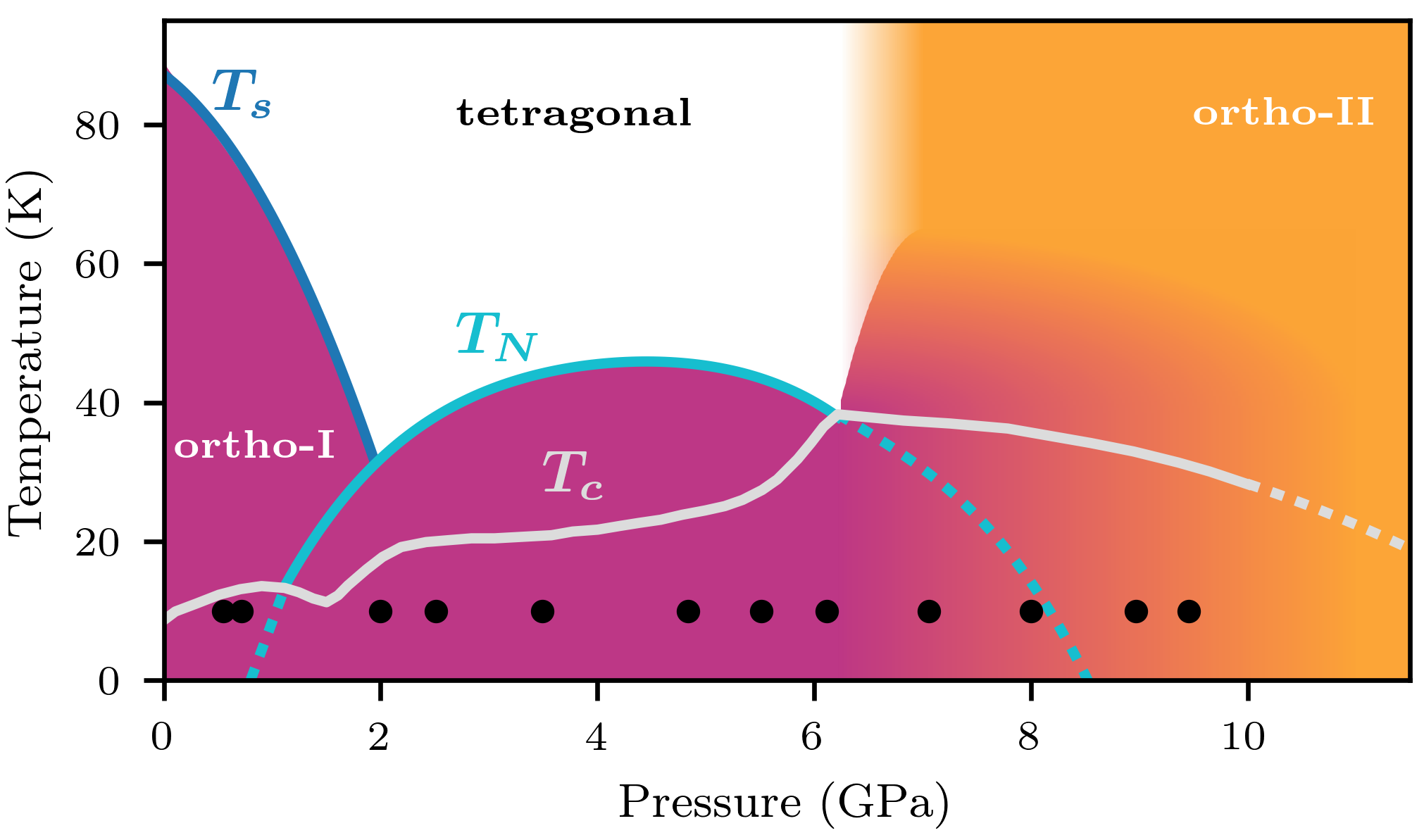} 
\caption{Schematic P-T phase diagram of FeSe \cite{Sun2016,Svitlyk2017}. Tetragonal (white), ortho-I (purple), and ortho-II (orange) structures are described in introduction. $T_s$, $T_N$ and $T_c$ are the structural, magnetic and superconducting transitions, with dashed lines represent extrapolations. P-T measurement points for XES/XAS are indicated with circles.}
\label{Fig:phase_diag}
\end{figure}

FeSe is the simplest form of the Fe superconductors (FeSC), yet one of the more fascinating members, as seen in its many overlapping phases and their apparent correlation with the superconducting critical temperature $T_c$ (Fig.~\ref{Fig:phase_diag}). 
%---JP---
However in spite of intensive research over the past years, the physics of FeSC remains still poorly understood because of the intricate coupling between structural, magnetic and electronic degrees of freedom. 
%-------- 
Below 6--7~GPa, FeSe has a PbO-type tetragonal structure ($P4/nmm$) at ambient temperature which is slightly distorted to an orthorhombic structure ($Cmma$) \cite{Margadonna2008} below the structural and magnetic transition temperatures ($T_s$ and $T_N$). The low-temperature orthorhombic phase, denoted here as ``ortho-I'', is characterized by nematic order with two distinct origins: below $T_s$, it is a second-order structural transition driven by orbital order \cite{Baek2015,Massat2016}; while below $T_N$, it is a simultaneous first-order magneto-structural transition driven by stripe-type spin fluctuations \cite{Wang2016,Wang2016PRL,Kothapalli2016}. Muon spin relaxation ($\mu$SR) \cite{Bendele1,Bendele2,KhasanovRotation} and nuclear forward scattering \cite{Kothapalli2016} find anti-ferromagnetic (AFM) order with a small magnetic moment in the 0.8--2.4~GPa range below T$_N$, although the exact type is unknown. Transport measurements show that this AFM order exists to at least 6.3~GPa \cite{Sun2016}. Above 6--7~GPa, a structural transition is observed \cite{Svitlyk2017,Garbarino2009,Margadonna2009,Kumar2010} to a MnP-type orthorhombic phase ($Pnma$ or equivalently $Pbnm$), denoted here as ``ortho-II'', with the Fe ion site symmetry changing from tetrahedral ($T_d$) to octahedral ($O_h$) coordination. This transition was observed with $^{57}$Fe M\"{o}ssbauer spectroscopy above 7~GPa as an additional component in the spectrum, however no magnetic hyperfine splitting is observed at 4.2~K 
\cite{Medvedev2009}. 
The superconducting critical temperature in FeSe shows drastic changes with pressure. After an initial increase, $T_c$ dips at the onset of AFM order, which is attributed to a reduction in the density of states due to a reconstruction of the Fermi surface \cite{Terashima2016,Terashima2016a}. Increasing the pressure further enhances both the magnetic and superconducting orders, with $T_N$ reaching a maximum of 45--55~K \cite{Bendele2,Sun2016} at $\sim$ 4.2~GPa and $T_c$ plateauing around 20~K below this pressure \cite{Sun2016,Miyoshi2014}. Above this pressure magnetic order decreases and coincides with a sharp rise in $T_c$ to a maximum around 37~K at $\sim$ 6.2~GPa \cite{Braithwaite2009,Margadonna2009,Garbarino2009,Medvedev2009,Sun2016}. After this maximum $T_c$ has a discontinuous change from a positive to negative slope, coinciding with the transition to the ortho-II phase \cite{Svitlyk2017}. An AFM region in the middle of a superconducting ``dome'' is quite unique, and the phase diagram shows their intimate connection as two $T_c$ anomalies where the two intersect. 

%---JP---
This macroscopic complexity reflects the strong coupling between charge, spin and structure which in turn affects the superconducting properties. It is therefore of great importance to be able to probe these different degrees of freedom at the local level while exploring the phase diagram in the vicinity of the superconducting regions. 
%bwl - removed microscopic from before local level
%-------
%The low pressure anomaly in $T_c$ evolution around 1~GPa has been thoroughly studied, however the high pressure one at 5--7~GPa has received relatively little attention, despite that fact that it is interesting due to the ortho-I $\rightarrow$ ortho-II transition in the same region. 
%
X-ray spectroscopy in the hard x-ray range is well-suited to that aim while being compatible with high pressure conditions \cite{Rueff2010}. In particular, K$\beta$ ($3p \rightarrow 1s$ transition) x-ray emission spectroscopy (XES) and x-ray absorption spectroscopy (XAS) at the K-edge are well-established probes of the local electronic, magnetic, and structural properties. As a primarily atomic probe, XES can access the local magnetic moment of a selected atom regardless of the magnetic order, while XAS provides a view of the local electronic and structural properties. In recent works \cite{Kumar2011,Chen2011}, \citeauthor{Kumar2011} and \citeauthor{Chen2011} have used XES to probe Fe magnetism in FeSe under high pressure. Their results show a decrease of the Fe spin state with a discontinuous change in slope near the ortho-I $\rightarrow$ ortho-II transition. This is interpreted as a high-spin (HS) to low-spin (LS) transition \cite{Kumar2011} or a state with a smaller magnetic moment \cite{Chen2011}. However, as demonstrated in this Letter, the HS state is excluded at low pressures. Furthermore, their experiments were performed with polycrystalline samples including the hexagonal polymorph of FeSe (11\% in the sample of \citeauthor{Chen2011}) and at room temperature or only up to 8~GPa at 8~K for \citeauthor{Kumar2011}. In this work, we perform a high-pressure study of FeSe single crystals at 10~K until 9.5~GPa using Fe-K$\beta$ XES and XAS at the Fe K-edge. 
%---JP---
Our results demonstrate a previously unobserved LS $\rightarrow$ HS transition in the same pressure region as the ortho-I $\rightarrow$ ortho-II transition and the maximum of $T_{c}$. The spin state transition coincides with the structural transition and $T_c$ slope change thus demonstrating the strong interplay between electronic and lattice degrees of freedom, adding to the remarkable properties of FeSe, and more generally to Fe superconductors. It is also to our knowledge the first LS to HS transition ever reported under pressure in a $3d$ metal compound. This challenges the classical crystal-field derived picture of spin transitions in these materials. We assign the low-spin state here to the atypical orbital-selective ``Mottness'' of the $3d$ electrons in Fe superconductors.  
%------

The XES/XAS experiment was performed on the GALAXIES beamline \cite{Rueff2015} at the SOLEIL synchrotron facility. We used high purity FeSe single crystals from the Institut N\'{e}el (Grenoble) grown by chemical vapor transport \cite{Karlsson2015}. Pressure was applied using a membrane-driven diamond anvil cell equipped with 1.2~mm thick diamonds with 300~$\mu$m culets. Several FeSe single crystals were loaded in a 150~$\mu$m hole of a CuBe gasket, along with ruby chips for in-situ pressure measurement \cite{RubyRubyRubyRubyyyy} and silicone oil as a pressure transmitting medium. XES/XAS were measured with the spectrometer in a transmission geometry using a 1-meter radius spherically bent Ge(620) crystal analyzer and an avalanche photodiode detector arranged in the Rowland circle geometry. The total energy resolution at the Fe~K$\beta$ line ($\sim$7057~eV, $\theta_B=79^{\circ}$) was 1.2~eV FWHM. The XES spectra were measured with 10~keV incident x-rays. The XAS spectra was measured using the partial-fluorescence yield (PFY) method with the spectrometer fixed to the Fe~K$\beta$ line. This technique leads to an intrinsic sharpening effect due to the shallower $3p$ core hole left in the final state with respect to the deeper $1s$ level \cite{Rueff2010}.
   
\begin{figure}[htbp]
\includegraphics[width=3.375in]{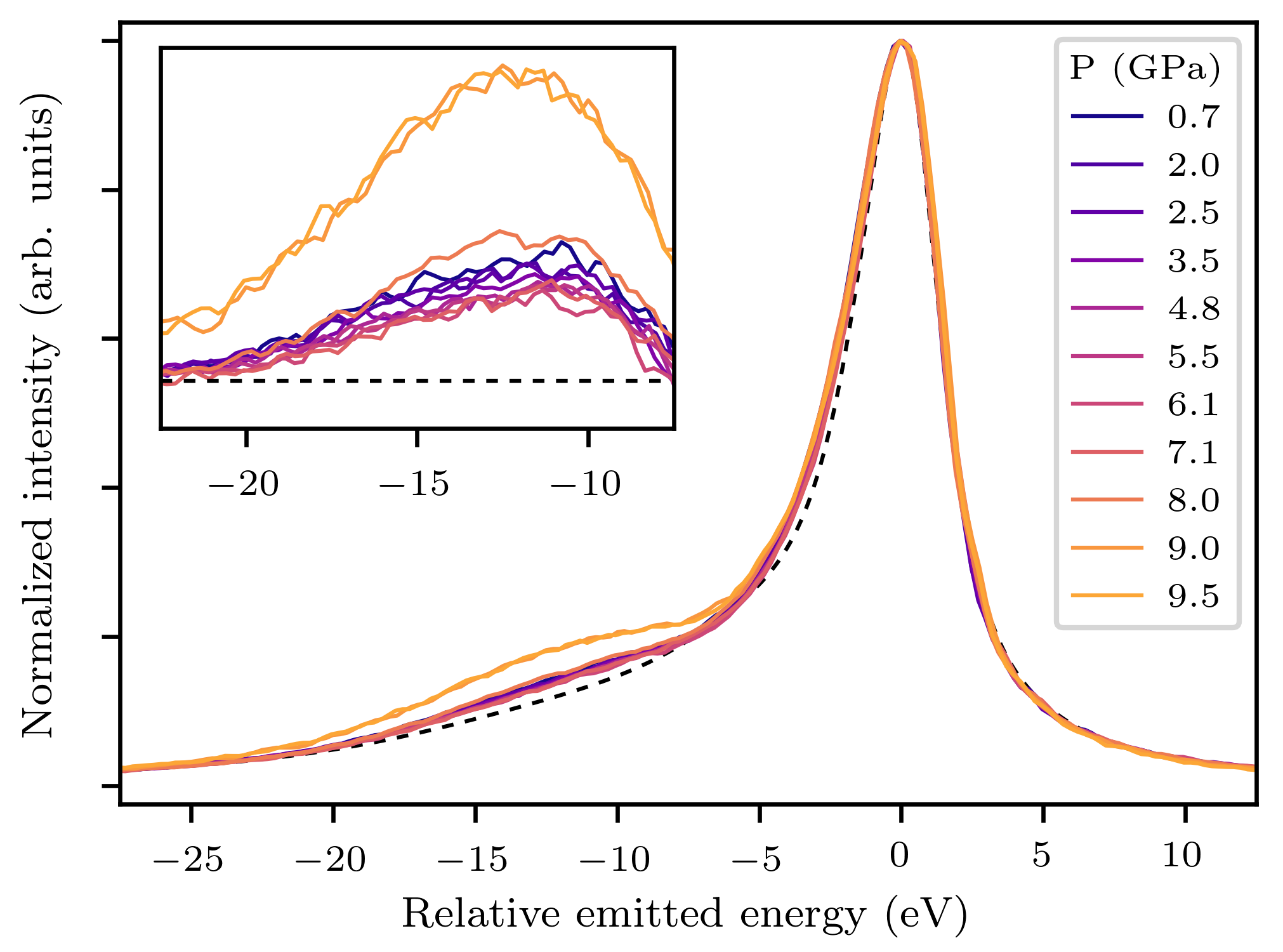}
\caption{Fe~K$\beta$ x-ray emission in FeSe as a function of pressure measured at 10~K. The inset is a zoom of the satellite region showing the difference with the FeS$_2$ zero-spin reference \cite{Lin2008}, shown as a black dashed line in the main plot.}
\label{Fig:XES}
\end{figure}

\begin{figure}[htbp]
\includegraphics[width=3.375in]{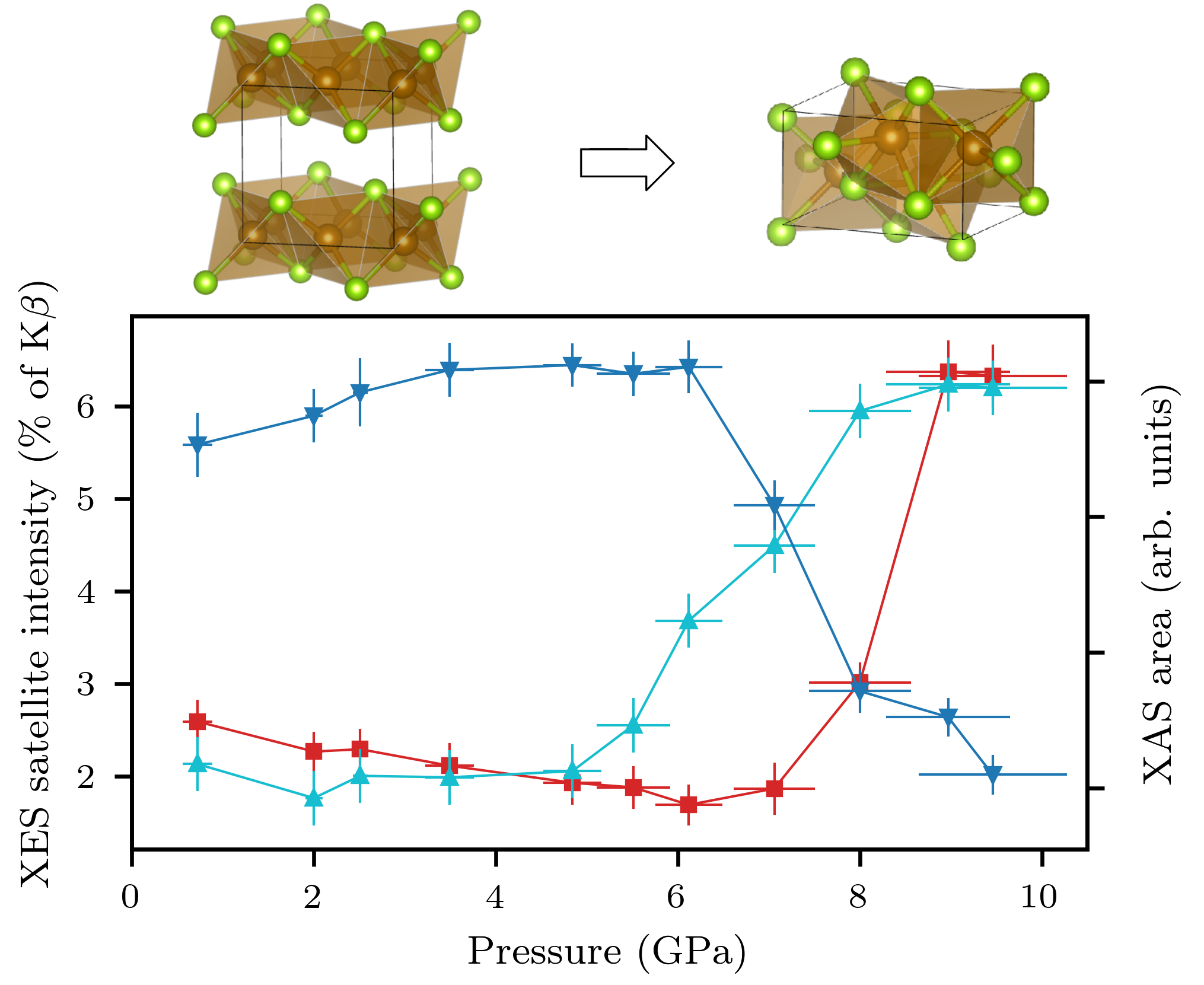}
\caption{(Top) Schematic of ortho-I to ortho-II structural transition. (Bottom) Pressure dependence of XES and XAS spectra at 10~K. (Left scale) Intensity of K$\beta$ satellite shown as red squares. (Right scale) Area of the Fe pre-edge feature A shown as blue inverted triangles and area between XAS features C and D shown as cyan triangles (scaled and offset).}
\label{Fig:IAD}
\end{figure}

Our XES spectra measured at 10~K are shown in Fig.~\ref{Fig:XES} for increasing pressure up to 9.5~GPa. The spectra are aligned to the main peak at 7057~eV and normalized to the maximum intensity. There is a weak satellite located around -12~eV from the main line, which is well-established as a signature of the local magnetic moment \cite{Peng1994}. The satellite can be seen more clearly by subtracting the zero-spin reference \cite{Lin2008}, FeS$_2$, as shown in the inset of Fig.~\ref{Fig:XES}. The fitted intensity of the satellite from the difference spectra is shown in Fig.~\ref{Fig:IAD} as red squares. FeSe exhibits a LS state with a gradually decreasing magnetic moment until 6--7~GPa. The magnetic moment starts to increase between 7 and 8~GPa and then jumps between 8 and 9~GPa to a HS state which plateaus around 9.5~GPa. The low-pressure behavior is consistent with previous results \cite{Kumar2011,Chen2011} and is understood to be due to band structure effects in the compressed lattice, however the sudden increase around 8~GPa was previously unobserved. The pressure region of the transformation of this LS $\rightarrow$ HS transition coincides with the ortho-I $\rightarrow$ ortho-II coexistence region at low temperature \cite{Svitlyk2017}. 

\begin{figure}[htbp]
\includegraphics[width=3.375in]{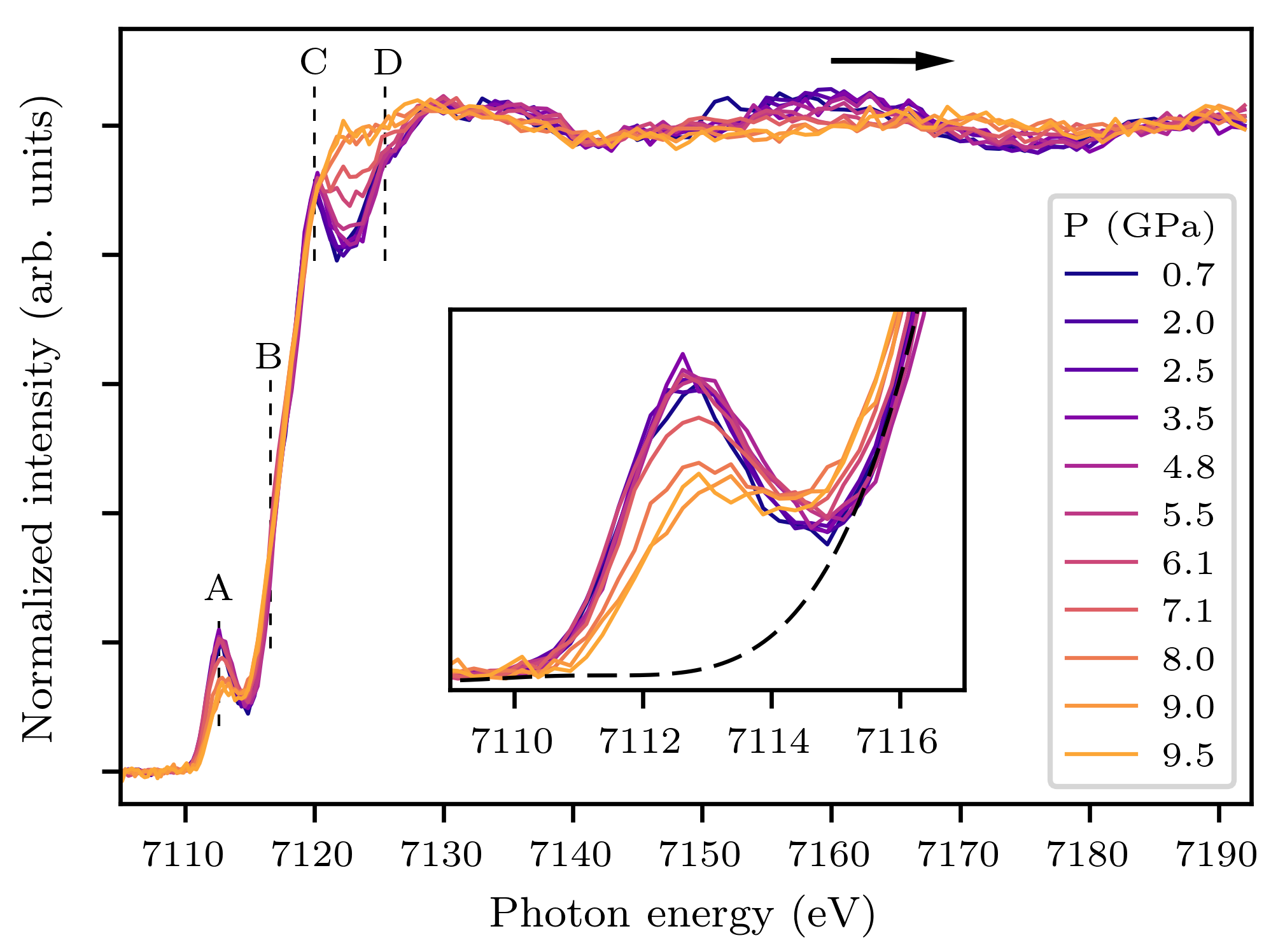}
\caption{Partial-fluorescence yield x-ray absorption at the Fe K-edge in FeSe as a function of pressure measured at 10~K. The inset is a zoom of the pre-edge region, with the Victoreen background (at 0 GPa) shown with dashed line.}
\label{Fig:XAS}
\end{figure}

To gain insight on the Fe local properties, we also measured high-resolution XAS at each pressure point after each XES measurement. The series of PFY-XAS spectra taken at the same pressures as XES is shown in Fig.~\ref{Fig:XAS}, where the inset emphasizes the pre-edge region which is sensitive to the $3d$ states. The spectra are normalized and flattened using the software \textsc{athena} \cite{ATHENA}. The pre-peak feature A in FeSe at low pressure is mainly due to the dipole transitions from Fe~$1s$ to Fe~$3d$-Se~$4p$ hybrid bands as expected in $T_d$ symmetry, with a smaller contribution from Fe quadrupole $1s \rightarrow 3d$ transitions \cite{Chen2011, Chen2012, Bendele3, Joseph2010}. Fe~$3d$-Se~$4p$ hybridization is sensitive to the local geometry, therefore the pre-peak can be used primarily to study the pressure evolution of the structure (and the spin state as discussed below). To clarify its pressure dependence, it was fitted by a Gaussian lineshape after subtraction of a Victoreen background to account for the rising edge. This pre-edge area is shown as blue inverted triangles in Fig.~\ref{Fig:IAD} using the right scale and the label ``XAS A''. As FeSe is compressed the pre-peak area slowly increases until 6~GPa, above which it decreases with a much sharper slope until our highest pressure point at 9.5~GPa. The decreased intensity is due to a reduced hybridization as the Fe-Se bond lengths increase by $\approx$5\% \cite{Margadonna2009} and the coordination of iron increases during the ortho-I ($T_d$) $\rightarrow$ ortho-II ($O_h$) transition \cite{Svitlyk2017}. The increased coordination is also supported by the change of the hyperfine splitting at high pressure \cite{Medvedev2009}. The remarkable parallel evolution of the magnetic (XES) and structural (XAS) properties shows the interplay between the electronic and lattice degrees of freedom. The fact that the structural transition precedes the magnetic transition suggests that it is not magnetically driven.

The absorption features B, C, and D can be assigned as Fe dipole $1s \rightarrow 4p$ transitions, where C and D also have significant contribution from Fe $1s$ to Fe~$4p$-Se~$3d$ hybrid band transitions \cite{Chen2011,Chen2012,Bendele3,Joseph2010}. The features at higher energies are dominated by multiple scattering of the photoelectron with the nearest neighbors. The feature around 7160~eV shows a clear shift (arrow in Fig.~\ref{Fig:XAS}) supporting the change of local structure we see with the pre-peak. A significant spectral change is observed in the region between features C and D, even before the structural and magnetic transitions. The change is shown in Fig.~\ref{Fig:IAD} (cyan triangles) by taking the area between C and D. The onset of the increased intensity corresponds remarkably well with the sudden increase in $T_c$ around 5~GPa \cite{Sun2016}. The trend is continuous until the ortho-II dominant phase at 9.5~GPa, which puts into question the possible connection between ortho-II and superconductivity since ortho-II is not believed to support superconductivity \cite{Svitlyk2017} and a negative d$\rho$/dT suggests it is semiconductor \cite{Sun2016,Medvedev2009}.

To further understand the structural and electronic transitions, the XAS spectra were simulated using the \textsc{fdmnes} code \cite{Guda2015}. A cluster radius of 12~\AA{} (10~\AA{}) for the low-pressure (high-pressure) reached full convergence. Spin-orbit coupling was included by relativistic effect corrections and the Fermi level was set self-consistently \cite{Bunau2009}. The crystal structures were taken from Ref.~\cite{Margadonna2008} for the low-pressure ortho-I phase and from Ref.~\cite{Kumar2010} for the high-pressure ortho-II phase. To simulate the pressure-induced spin-state transition, the calculations in both phases were carried out using either a LS or a HS ground state configuration. 
%---JP---- 
The energy scales of the calculations have been stretched 10\% to account for systematic energy shifts due to the real part of the self energy function used to describe the excited state in these calculations \cite{Rehr2017}. 
%----

\begin{figure}[htbp]
\includegraphics[width=3.375in]{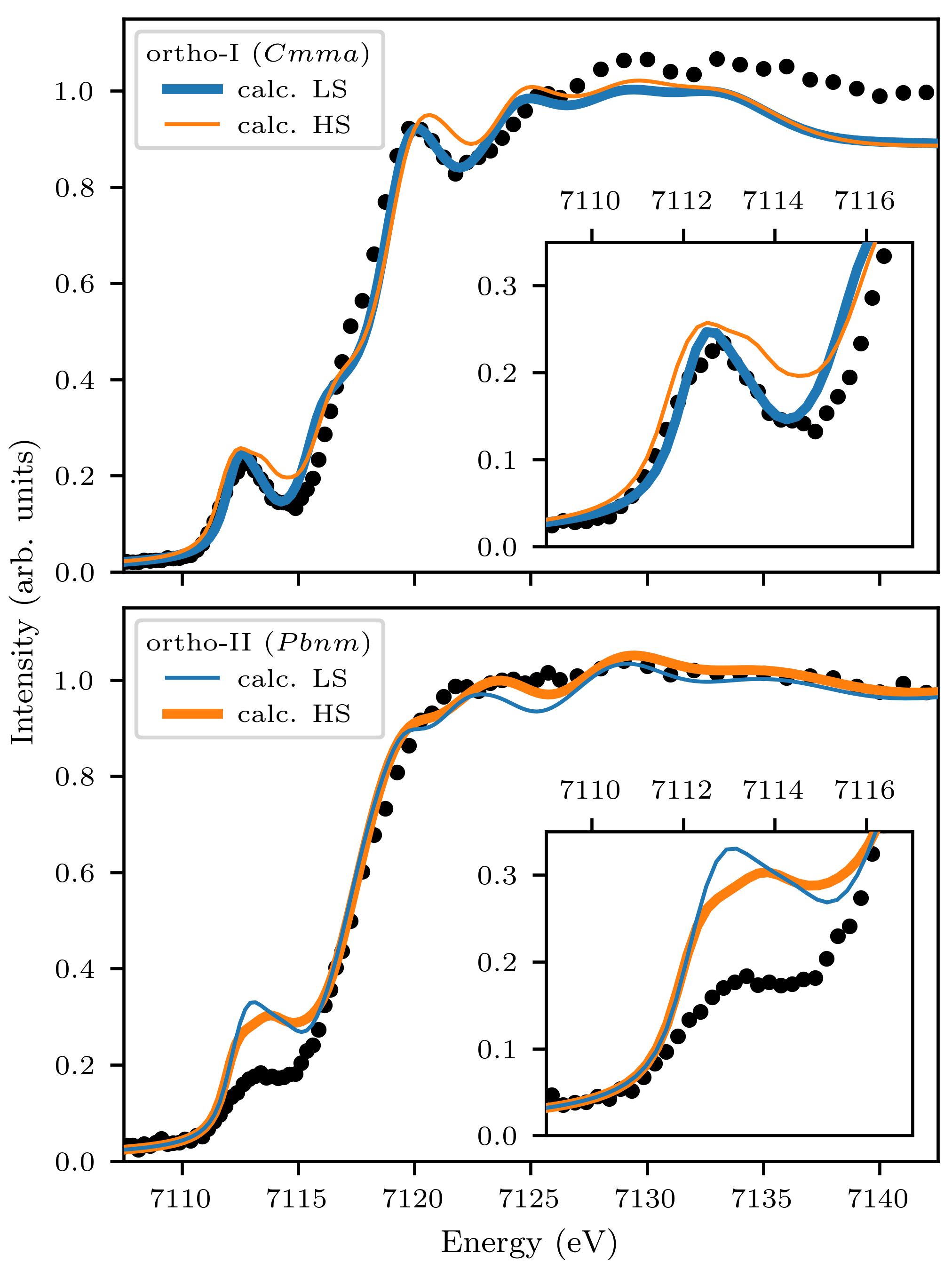}
\caption{Calculated (lines) XAS spectra in LS and HS configurations at the Fe K-edge in FeSe compared to the experimental spectra (circles) at 10~K. (Top panel) The ortho-I spectrum at 0.7~GPa. (Bottom panel) The ortho-II spectrum at 9.0~GPa. The insets are zooms of the pre-edge region. The spin state which agrees best with the experimental pre-peak region is drawn in bold.}
\label{Fig:calc}
\end{figure}

The results are shown in Fig.~\ref{Fig:calc} along with the experimental spectra measured at 0.7~GPa and 9.0~GPa. We do not expect to yield an accurate description of the $d$ electronic structure without including correlations --- more accurate approaches show that many-body effects lead to a renormalization of the $d$ bands and changes of the $d$ density of states close to the Fermi edge \cite{Aichhorn2010} --- but the K-edge should be well-described since it mostly connects to a single particle picture. In FeSe, we find that the calculations for both the low-pressure and high-pressure phases compare well to the experimental spectra in the pre-edge and near-edge regions. We notice however that the intensity of the pre-edge for the high pressure phase is overestimated.
%---JP---
The pre-edge intensity is strongly dependent on minor distortions of the metal site which favor or disfavor dipolar transitions to ligand $p$ states in addition to the weaker quadrupolar transition to the $d$ states. On the contrary, the pre-edge shape is mostly related to the orbitals' energy splitting and spin state, therefore it is a much better fingerprint of the local $3d$ electronic properties.
%-------
Indeed, as shown in the insets of Fig.~\ref{Fig:calc}, the pre-edge shape matches very well with a high-spin (low-spin) configuration for the high-pressure (low-pressure) phase. The less-pronounced pre-peak in ortho-II is consistent with high-spin Fe\textsuperscript{II} in an octahedral environment \cite{Westre1997}. This agrees with our XES results that there is a transition to a high-spin state above 7~GPa at 10~K in FeSe.

The main outcome of our study is the original observation of a high-spin state above 7 GPa as FeSe adopts the ortho-II phase. This follows the initial decrease of the local moment in the ortho-I phase which has been reported elsewhere \cite{Chen2011,Kumar2011} and confirmed here. Both XAS and XES indicate that the magnetic surge at high pressure is consistent with a pressure-induced LS to HS transition. The electronic change is likely resulting from the change of Fe site symmetry: in the ortho-I phase, Fe occupies a tetrahedral ($T_d$) site that becomes increasingly distorted with pressure whereas Fe sits in an octahedral ($O_h$) site in the ortho-II phase \cite{Svitlyk2017,Margadonna2009,Garbarino2009}. At first glance, the spin state of FeSe is surprising. According to ligand field theory, one would expect Fe to be high-spin in the ortho-I phase since tetrahedral coordination normally favors a high-spin state. However, the iron superconductor's electronic structure was also shown to be controlled by the correlated and specific ``Mottness'' of $d$ electrons with band-dependent correlations yielding $e_g$ states less correlated than $t_{2g}$ states \cite{Medici2014}. We suggest that the $T_d$ low-spin state is stabilized because it minimizes the on-site electron correlations by filling preferentially the $e_g$ bands. At high pressure, correlation becomes less effective as the bandwidth broadens which eventually allows a high-spin configuration as the Fe symmetry turns $O_h$. 
%---JP----
In line with recent band structure calculations \cite{Parker2017}, we can further argue that the exceptional stability of the magnetism at high pressure in FeSe could be aided by the increased 3-dimensional character of the compressed lattice and change of the Fermi surface.
%---------

%We report in this Letter a low-spin to high-spin transition in compressed FeSe, the first ever reported in a $3d$ metal compound to our knowledge. The spin state transition coincides with the structural transition and $T_c$ slope change thus demonstrating the strong interplay between electronic and lattice degrees of freedom in FeSe.

%We lack here the knowledge of the magnetic order as x-ray spectroscopy are local probes.  We performed neutron powder diffraction until $\sim$12~GPa, yet did not detect a significant amount of the ortho-II phase in order to search for magnetic order (see Supplemental Material). This calls for neutron powder diffraction at even higher pressures, which was shown to be feasible to at least 20~GPa at low temperatures \cite{Klotz2016}. 

\begin{acknowledgments}
We acknowledge SOLEIL for provision of synchrotron radiation facilities (proposal 20151119) and help from the high pressure laboratory for cell loading. We acknowledge P. Strobel and S. Karlsson from Institut N\'{e}el, Grenoble, France, for their help in the growth of the FeSe crystals using the CVT method.  B.W.L acknowledges financial support from the French state funds managed by the ANR within the ``Investissements d'Avenir'' programme under reference ANR-11-IDEX-0004-02, and within the framework of the Cluster of Excellence MATISSE led by Sorbonne Universit\'{e} and from the LLB/SOLEIL PhD fellowship program. 
%---JP----
%remove this part if length too long
%Sample synthesis and preparation was done by P.T. The XES/XAS experiment was performed by B.W.L, J.P.R, P.T, V.B, and J.M.A. The XES/XAS data treatment was done by B.W.L and J.P.R. The FDMNES simulations were performed by V.B, J.P.R, and B.W.L. The paper was written by B.W.L and J.P.R with the aid of the other co-authors.
%-----
\end{acknowledgments}
%We acknowledge ILL for provision of neutron facilities (proposal 5-31-2517 \cite{ILLexp}).
%The NPD experiment was performed by B.W.L, V.B, P.T, S.K, P.R, M.R, and T.H. The NPD data was treated by B.W.L, P.T, and V.B. 

\bibliography{main}
\end{document}